\documentclass[pra,twocolumn,preprintnumbers,amsmath,amssymb,superscriptaddress]{revtex4-1}

\usepackage[ansinew]{inputenc}
\usepackage[T1]{fontenc}
\usepackage{ae,aecompl}
\usepackage[english]{babel}
\usepackage[dvips]{graphicx}
\usepackage{dsfont}
\usepackage{amsmath}
\usepackage{enumerate,amsthm,amsmath,amssymb,color,
float}
\usepackage{pifont}
\usepackage{eufrak}

\newcommand{\ketbra}[2]{\mbox{$|#1\rangle\langle #2|$}}
\newcommand{\op}[1]{\mbox{\boldmath $\hat{#1}$}}

\newcommand{\ket}[1]{\vert#1\rangle}
\newcommand{\bra}[1]{\langle#1\vert}

\begin{document}

\title{Improving the maximum transmission distance of continuous-variable quantum key distribution using a noiseless amplifier}
\author{R\'emi Blandino}
\email{remi.blandino@institutoptique.fr}
\affiliation{Laboratoire Charles Fabry, Institut d'Optique, 
CNRS, Universit\'e Paris-Sud, Campus Polytechnique, RD 128, 91127 Palaiseau cedex, France}

\author{Anthony Leverrier}
\affiliation{Institute for Theoretical Physics, ETH Zurich, 8093 Zurich, Switzerland}

\author{Marco Barbieri}
\altaffiliation{Current address: Clarendon Laboratory, Department of Physics, University of Oxford, OX1 3PU, United Kingdom}
\affiliation{Laboratoire Charles Fabry, Institut d'Optique, 
CNRS, Universit\'e Paris-Sud, Campus Polytechnique, RD 128, 91127 Palaiseau cedex, France}
\author{Jean Etesse}
\affiliation{Laboratoire Charles Fabry, Institut d'Optique, 
CNRS, Universit\'e Paris-Sud, Campus Polytechnique, RD 128, 91127 Palaiseau cedex, France}
\author{Philippe Grangier}
\affiliation{Laboratoire Charles Fabry, Institut d'Optique, 
CNRS, Universit\'e Paris-Sud, Campus Polytechnique, RD 128, 91127 Palaiseau cedex, France}
\author{Rosa Tualle-Brouri}
\affiliation{Laboratoire Charles Fabry, Institut d'Optique, 
CNRS, Universit\'e Paris-Sud, Campus Polytechnique, RD 128, 91127 Palaiseau cedex, France}
\affiliation{Institut Universitaire de France, 103 boulevard St. Michel, 75005, Paris, France}
~\\
\begin{abstract}
We show that the maximum transmission distance of continuous-variable quantum key distribution in presence of a Gaussian noisy lossy channel can be arbitrarily increased using a linear noiseless amplifier. We explicitly consider a  protocol using amplitude and phase modulated coherent states with reverse reconciliation. We find that a noiseless amplifier with amplitude gain $g$ can increase the maximum admissible losses by a factor $g^{-2}$.
\end{abstract}
\maketitle


\section{Introduction}
Cryptography is certainly one of the most advanced applications of quantum technologies. 
Within this field, the most studied primitive is quantum key distribution (QKD), which is the art of distributing a secret key to two distant parties, Alice and Bob, in an untrusted environment controlled by an adversary, Eve \cite{SBC08}.
The security of QKD lies on the idea that an adversary trying to acquire some information about the secret key will necessarily introduce some noise in the quantum communication between Alice and Bob. A consequence of this idea is that if the quantum channel is too lossy or noisy, then it cannot be used to distill a secret key. This limits the maximum transmission distance between the legitimate parties. Developing QKD protocols resistant to losses and noise is therefore of great practical importance. 

Among QKD protocols, those encoding information in the amplitude and phase of coherent states \cite{grosshans_continuous_2002} have the advantage of only requiring off-the-shelf telecom components, as well as being compatible with wavelength-division multiplexing \cite{qi2010feasibility}, making an interesting solution for robust implementations \cite{cerf_quantum_2007,weedbrook_gaussian_2012}. 

On the theoretical side, these continuous-variable (CV) protocols have been proven secure against arbitrary attacks provided that they are secure against collective attacks \cite{RC09}. This latter condition is in particular met for all CV protocols without postselection for which Gaussian attacks are known to be optimal within collective attacks \cite{GC06,NGA06,PBL08,leverrier_simple_2010}.

Protocols with postselection on the other hand \cite{SRL02,LKL04}, where Alice and Bob only use part of their data to extract a secret key, can increase the robustness of QKD to losses and noise but at the price of more involved security proofs. In particular, their security is only established against Gaussian attacks \cite{HL06,HL07}, or when an active symmetrization of the classical data is applied \cite{PhysRevA.85.022339}.

In this paper, we consider the use of a noiseless linear amplifier (NLA) \cite{ralph_nondeterministic_2008,ferreyrol_implementation_2010,ferreyrol_experimental_2011,barbieri_nondeterministic_2011,zavatta_high-fidelity_2011,xiang_heralded_2010} on the detection stage as a way to increase the robustness of CV QKD protocols against losses and noise. First, it should be noted that while amplifiers can effectively recover classical signals, they only offer limited advantages when working on quantum signals, as amplification is bound to preserve the original signal to noise ratio (SNR) \cite{caves_quantum_1982,levenson_reduction_1993,ferreyrol_experimental_2011}. This implies that ordinary linear amplifiers, as those realized by optical parametric processes \cite{loudon_quantum_2000}, can only find limited applications in the context of QKD \cite{fossier_improvement_2009}.



On the other hand, a \emph{probabilistic} NLA can in principle amplify the amplitude of a coherent state while retaining the initial level of noise \cite{ralph_nondeterministic_2008}. Thus, when only considering its successful runs, the NLA can compensate the effect of losses and could therefore be useful for quantum communication \cite{ralph_quantum_2011}, and to establish the nonlocal nature of quantum correlations thanks to a loophole-free Bell test \cite{PhysRevA.85.042116}.
The availability of such a device has stimulated intense experimental activity over the past years, demonstrating the implementation of approximated  versions \cite{ferreyrol_implementation_2010,zavatta_high-fidelity_2011,xiang_heralded_2010,ferreyrol_experimental_2011,barbieri_nondeterministic_2011}, which have provided solid proof-of-principle. 

The question arises if these more sophisticated devices can deliver a compensation of losses with a success rate such that it may represent a useful tool for quantum cryptography. Here we address this problem, by investigating the advantages and limitations of the most general NLA device, without making assumptions on the particular realization. 

We find a regime in which the NLA leads to an improvement of the maximum transmission distance attainable on a noisy and lossy Gaussian channel. Because of the non-deterministic nature of the NLA, the security proofs considered here are similar to those concerning protocols with postselection, that is, they hold against Gaussian attacks, or collective attacks provided an additional symmetrization of the classical data is performed. 


\section{Description of the GG02 protocol}

We consider explicitly the case for the most common protocol for continuous-variable QKD, designed by Grosshans and Grangier (GG02) \cite{grosshans_continuous_2002}. In its prepare-and-measure (PM) version, Alice encodes information in the quadratures of coherent states which are then sent to Bob through the untrusted quantum channel. Alice chooses  her preparation $\ket{\alpha{=}x_\mathrm{A}{+}ip_\mathrm{A}}$ from a Gaussian distribution for the two quadratures having zero mean and variance $V_\mathrm{A}$. Bob randomly decides whether to measure the 
$\hat{x}$ or the $\hat{p}$ quadrature, using homodyne detection. Alice and Bob finally extract a secret key from the correlated data by performing classical data manipulation and authenticated classical communication.  This protocol offers a simple experimental implementation \cite{grosshans_quantum_2003,LBG07,fossier_field_2009,jouguet2012field} and is secure against finite-size collective attacks \cite{LGG10} as well as arbitrary attacks in the asymptotic limit of arbitrary long keys \cite{RC09}. 

This protocol can be reformulated in an entanglement-based version (EB), in terms of entanglement distribution between Alice and Bob \cite{grosshans_virtual_2003}: the two parties initially share a two-mode squeezed vacuum state $\ket{\lambda}{=}\sqrt{1-\lambda^2}\sum_{n{=}0}^{\infty}\lambda^n\ket{n}\ket{n}$, with $\lambda{<}1$. Alice  performs an \mbox{heterodyne} measurement on her mode, which projects the other mode on a coherent state. The outcome of Alice's measurement is random, but with a probability distribution depending on $\lambda$. 

Although the EB version does not correspond to the actual implementation, it is fully equivalent to the PM version from a security point of view, and it provides a more powerful description for establishing security proofs against collective attacks through the covariance matrix $\gamma_\mathrm{AB}$ of the state shared by Alice and Bob before their respective measurements. In the case of a Gaussian channel with transmittance $T$, and input equivalent excess noise $\epsilon$ \cite{weedbrook_gaussian_2012}:
\begin{align}
\gamma_\mathrm{AB}= \left(
\begin{array}{cc}
	V(\lambda) \mathbb{I} &\sqrt{T(V(\lambda)^2-1)} \mathbb{Z} \\
	\sqrt{T(V(\lambda)^2-1)} \mathbb{Z} & T(V(\lambda)+B+\epsilon) \mathbb{I}
	\end{array} \right)
	\label{CM}
\end{align}
where $\mathbb{I}{=}\operatorname{diag}(1,1)$ and $\mathbb{Z}{=}\operatorname{diag}(1,{-}1)$, $V(\lambda){=}\frac{1{+}\lambda^2}{1{-}\lambda^2}$ is the variance of the thermal state $\operatorname{Tr}_\mathrm{A}\ket{\lambda}\bra{\lambda}$ related to the modulation variance by $V_\mathrm{A}{=}V{-}1$, and $B{=}\frac{1-T}{T}$ is the input equivalent noise due to losses.

This matrix contains all the information needed to establish the secret key rate for collective attacks \cite{LBG07}:
\begin{align}
\Delta I(\lambda,T,\epsilon,\beta){=}\beta I_\mathrm{AB}(\lambda,T,\epsilon)-\chi_\mathrm{BE}(\lambda,T,\epsilon),
\label{secretKey}
\end{align}
where $I_\mathrm{AB}{=}\frac{1}{2}\log_{2}(\frac{V{+}B{+}\epsilon}{1{+}B{+}\epsilon})$ is the mutual information shared by Alice and Bob given by Shannon's theory \cite{shannon_mathematical_1948}, and $\chi_\mathrm{BE}$ is the Holevo bound for the mutual information shared by Eve and Bob (see Appendix \ref{appendixHolevo}). The reconciliation efficiency $\beta{<}1$ accounts for the fact that in practical implementations of this protocol, Alice and Bob do not have sufficient resources to reach the Shannon limit. Steady progress has been made in recent years on the problem of error correction for CV QKD \cite{bloch_ldpc-based_2006,LAB08,jouguet2012high} and today procedures based on modern error correcting techniques achieve $\beta \approx95 \%$ for a large range of SNR \cite{JKL11}.


\section{Equivalent channel and squeezing}
\label{equivalentChannel}

Let us now consider the use of a NLA in the GG02 protocol. As usual, we will perform the security analysis of the EB version. Here, we restrict ourselves to the case of a Gaussian quantum channel, that is Eve is limited to performing Gaussian attacks. Since the secure key rate of the protocol depends only on the covariance matrix of Alice and Bob $\gamma_\mathrm{AB}$, it is sufficient to compute it in presence of the NLA.

In this modified version of the protocol, Alice and Bob implement GG02 as usual but Bob adds a NLA to his detection stage, before his homodyne detection, which is here assumed to be perfect. Then, only the events corresponding to a successful amplification will be used to extract a secret key. This scheme is therefore very similar to protocols with postselection.


Since the output of the NLA remains in the Gaussian regime, we can look for equivalent parameters of an EPR state sent through a Gaussian noisy channel. Their derivation is explained in detail in Appendix \ref{appendixEffectiveParameters}, where it is shown that the covariance matrix $\gamma_\mathrm{AB}(\lambda,T,\epsilon,g)$ of the amplified state is equal to the covariance matrix $\gamma_\mathrm{AB}(\zeta,\eta,\epsilon^g,g{=}1)$ of an equivalent system with an EPR parameter $\zeta$, sent through a channel of transmittance $\eta$ and excess noise $\epsilon^g$, without using the NLA. Those effective parameters are given by:
\begin{equation} 
	\addtolength{\fboxsep}{4pt} 
				\begin{split} 
					\zeta&=\lambda  \sqrt{\frac{\left(g^2-1\right) \left(\epsilon-2\right) T-2}{\left(g^2-1\right)
   \epsilon T-2}}, \\
\eta&=\frac{g^2 T}{\left(g^2-1\right) T \left(\frac{1}{4} \left(g^2-1\right) \left(\epsilon-2\right) \epsilon T-\epsilon+1\right)+1} , \\
\epsilon^g&=\epsilon-\frac{1}{2} \left(g^2-1\right) \left(\epsilon-2\right) \epsilon T.
					\label{parametresEffectifsMain}
				\end{split}  
	\end{equation}

\begin{figure}[h]
\centering
\includegraphics[width=\columnwidth]{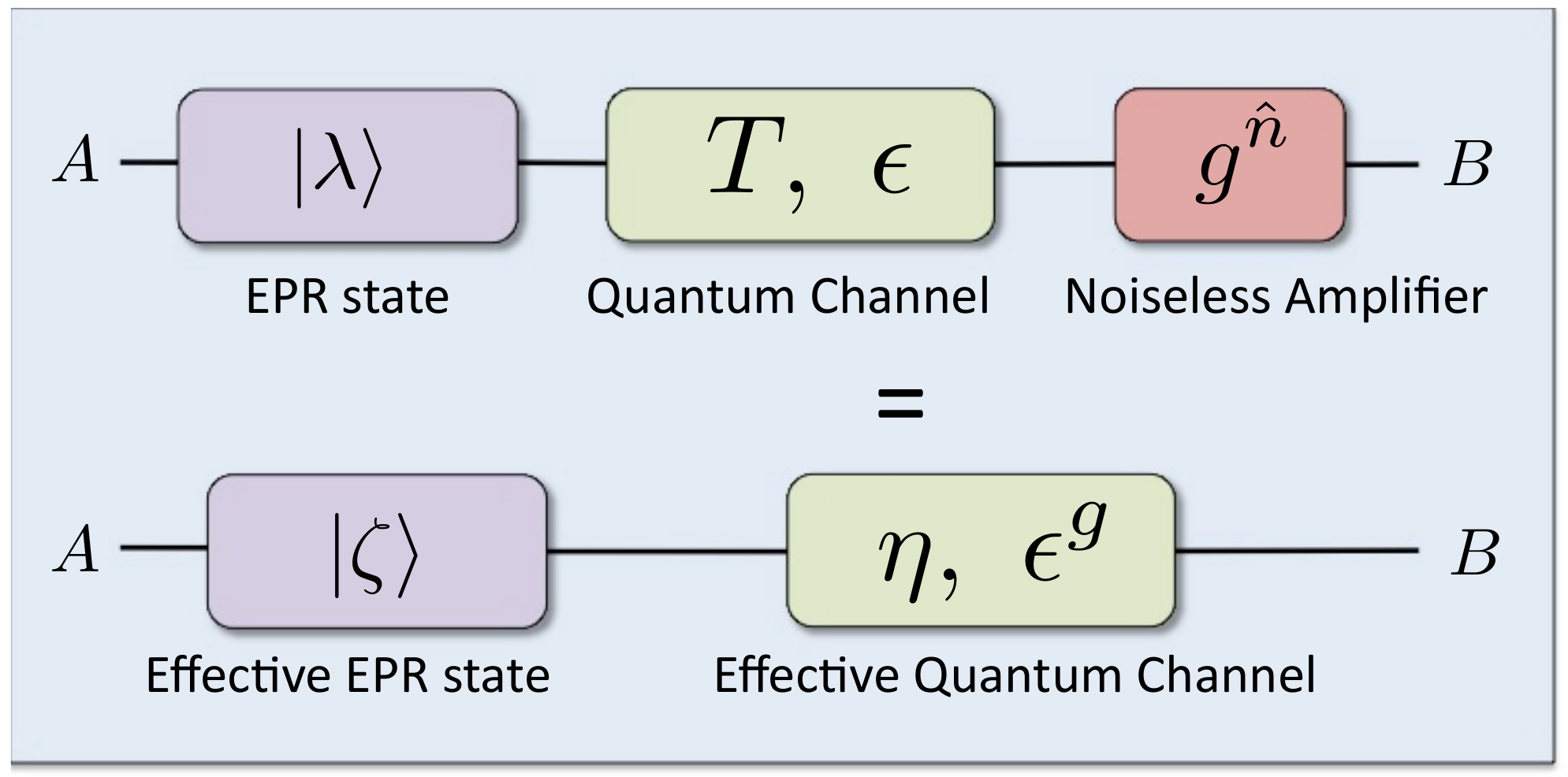} 
\caption{Equivalent channel and squeezing: a state $\ket{\lambda}$ sent through a Gaussian channel of transmittance $T$ and excess noise $\epsilon$, followed by a successful amplification, has the same Alice-Bob covariance matrix than a state $\ket{\zeta}$ sent through a Gaussian channel of transmittance $\eta$ and excess noise $\epsilon^g$, without the NLA.}
\label{schema}
\end{figure}

This identification easily provides the secret information $\Delta I ^g$ corresponding to the successful amplification, since Eq. \ref{secretKey} can be used with the effective parameters:
 \begin{align}
 \Delta I ^g (\lambda, T,\epsilon,\beta)= \Delta I(\zeta,\eta,\epsilon^g,\beta).
 \label{deltaMaxNLA}
 \end{align}
 
Those parameters can be interpreted as physical parameters of an equivalent system if they satisfy the physical meaning constraints $0{\leq}\zeta{<}1$, $0{\leq}\eta{\leq}1$, and $\epsilon^g{\geq}0$. Since $\lambda$ is a global factor in the expression of $\zeta$, the first condition is always satisfied if $\lambda$ is below a limit value:
\begin{align}
0\leq\zeta < 1 \Rightarrow 0<\lambda < \left(\sqrt{\frac{\left(g^2{-}1\right) \left(\epsilon{-}2\right) T{-}2}{\left(g^2{-}1\right)
   \epsilon T-2}}\right)^{{-}1}.
\end{align}
As $\eta$ and $\epsilon^g$ do not depend on $\lambda$, the parameter $\zeta$ can be considered as independent of those two parameters, keeping in mind that this simply sets the value of $\lambda$. 

\begin{figure}[H]
\centering
\includegraphics[width=6.5cm]{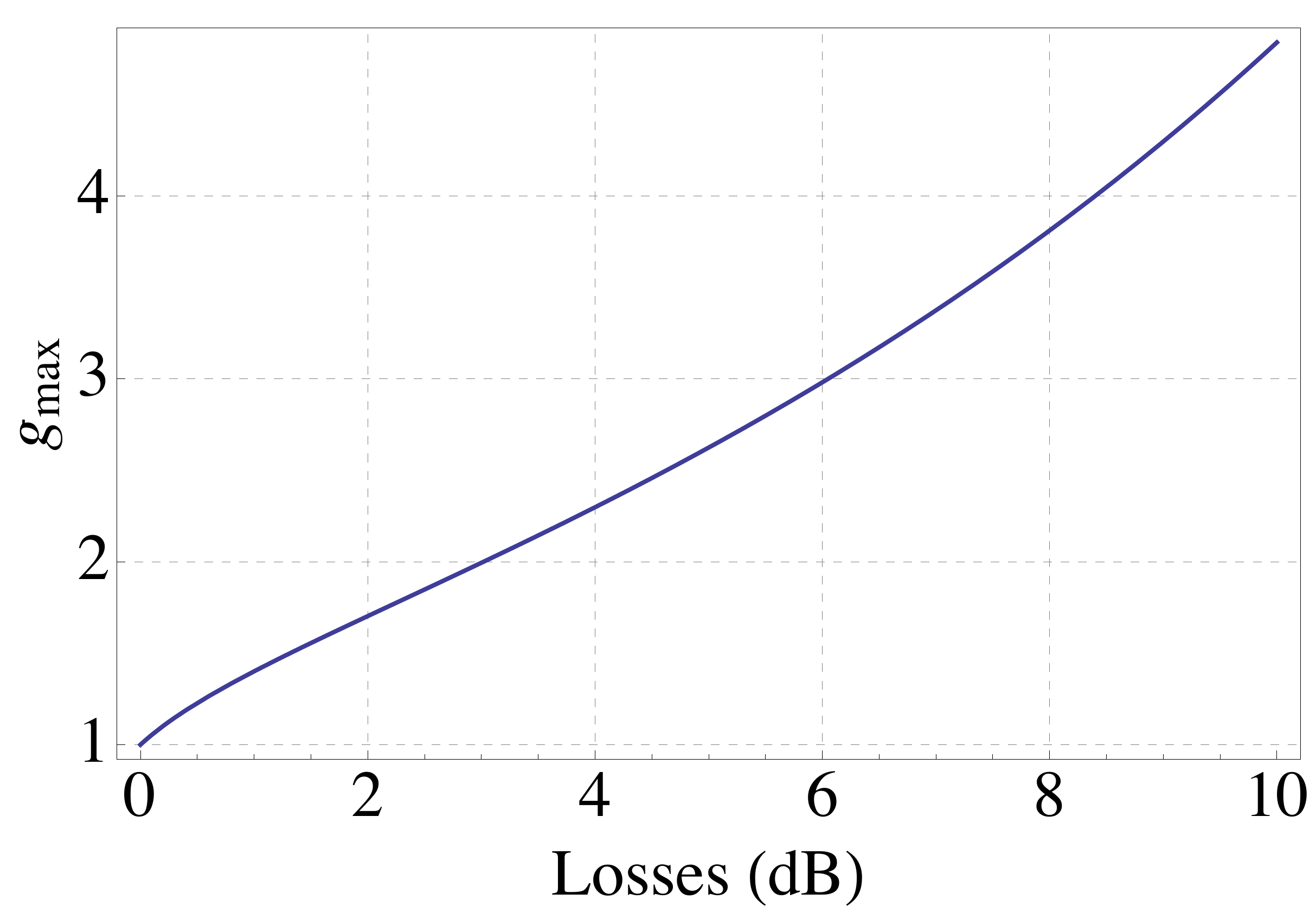} 
\caption{$g_\mathrm{max}(T,\epsilon)$ against the losses in dB. $\epsilon{=}0.2$.}
\label{gMax}
\end{figure}

The second and the third conditions are satisfied if the excess noise is smaller than 2, and if the gain is smaller than a maximum value given by Eq. \ref{gainMax}, and plotted on Fig. \ref{gMax}.

\section{Increase of the maximum transmission distance}
The analysis of the equivalent state allows us to compare the secret key rate obtained with and without an ideal NLA. The comparison must be performed for a given channel with fixed losses $T$ and excess noise $\epsilon$, as those parameters cannot be controlled by Alice or Bob. However, since the relevant quantity is the maximum secret key rate achievable over this channel, Alice is allowed to optimize her modulation variance $V_\mathrm{A}$ (or equivalently, the parameter $\lambda$) in order to maximize the secret key rate.

The secret key rate without the NLA is given by $\Delta I(\lambda,T,\epsilon,\beta)$ (Eq. \ref{secretKey}).
The secret key rate with the NLA $\Delta I_\mathrm{NLA}$ is obtained by multiplying the secret key rate for successful amplifications $\Delta I^g$ by the probability of success $P_\mathrm{suc}$. If the NLA has a sufficient dynamics to neglect distortions, we can assume that $P_\mathrm{suc}$ is constant.  This is a reasonable assumption if $\beta{<}1$, since in that case the optimal value of $V_\mathrm{A}$ is not infinite. The precise value of $P_\mathrm{suc}$ will depend on practical implementations, and is not important is our study, since it only acts as a scaling factor and does not change the fact that a negative secret key rate can become positive with a NLA.
Therefore, 
\begin{align}
\Delta I_\mathrm{NLA}=P_\mathrm{suc}\Delta I(\zeta,\eta,\epsilon^g).
\end{align}

In Appendix \ref{appendixProba}, we show that the probability of success for a NLA of gain $g$ is upper bounded by $1/g^2$. We can therefore use this bound, keeping in mind that
the relevant conclusion which can be taken is only whether the secret key rate is positive or not.
Both secret key rate with and without the NLA are computed using the formulae given in Appendix \ref{appendixHolevo}.

Since the expression of $\Delta I$ is relatively difficult to manipulate, we perform a series expansion at the first order in $T$, which corresponds to the strong losses regime (Appendix \ref{appendixDLHolevo}). The approximate secret key rate is given by Eq. \ref{dlHolevo}. Its expression gives us an intuition about two important behaviors: first, since $T$ appears inside the expansion and not only as a global factor, it explains why there can be a maximum transmission distance, or equivalently a value $T_\mathrm{lim}$ for which the secret key rate becomes null. Second, in this regime, the effect of the NLA is simply to replace the transmittance $T$ by $g^2 T$, the other physical parameters being the same. Hence, it is clear that the losses are reduced, which will increase the maximum distance of transmission.

Let us prove those statements more precisely. From Eq. \ref{dlHolevo}, we find an analytical value of $T_\mathrm{lim}$ when $g{=}1$ (\textit{i.e} without the NLA),
\begin{align}
&T_\mathrm{lim}=\frac{1}{\epsilon}2 \lambda ^{-\frac{4 \lambda ^4}{\epsilon  \left(\lambda ^2-1\right)^2}}
   e^{\frac{\lambda ^2 (2 \beta +\epsilon )-\epsilon }{\epsilon  \left(\lambda
   ^2-1\right)}}.
   \label{Tlim}
\end{align}
This expression clearly tends to 0 when $\epsilon$ tends to 0, which shows that there is no maximum transmission distance without excess noise. Interestingly, there is a maximum transmission distance as soon as the excess noise $\epsilon$ is non zero, even if the reconciliation efficiency $\beta$ equals 1. When $\beta$ decreases or when $\epsilon$ increases, this maximum transmission distance decreases. There is no limitation of the distance of transmission only when $\epsilon{=}0$, and in that case Eq. \ref{dlHolevo} takes a simple form:
\begin{align}
\Delta I_\mathrm{NLA}&{\simeq}\frac{1}{g^2} g^2 T \lambda ^2  \frac{ \left(1{-}\lambda ^2\right) (\beta {-}2 \log \lambda
   ){+}2 \log \lambda}{\left(\lambda ^2{-}1\right)^2 \log (2)}{\simeq}\Delta I.
   \label{tauxNoNoise}
\end{align}
 This shows that for strong losses without excess noise, the secret key rate using the NLA with the most optimistic probability of success is the same as the secret key rate without the NLA, and is always positive if $\lambda$ is optimized. $T_\mathrm{lim}$ can also be optimized (\textit{i.e.} minimized) by optimizing $\lambda$. Interestingly, the optimal value $\lambda_\mathrm{opt}$ depends only on $\beta$, as shown by Eq. \ref{betaOpt}.

The same calculation with a NLA of gain $g$ shows that:
\begin{align}
T_\mathrm{lim}^g=\frac{1}{g^2}T_\mathrm{lim}.
\end{align}
Therefore, the losses for which the secret key rate is zero are increased by:
\begin{align}
\Delta \mathfrak{L}=20 \log_{10} g \text{ dB}. 
\end{align}

Let us stress that this result does not depend on the probability of success of the NLA, which simply acts as a scaling factor for the secret key rate. Hence, even for a more realistic probability of success, the NLA increases the maximum distance of transmission in the same way.

\begin{figure}[t]
\centering
\includegraphics[width=\columnwidth]{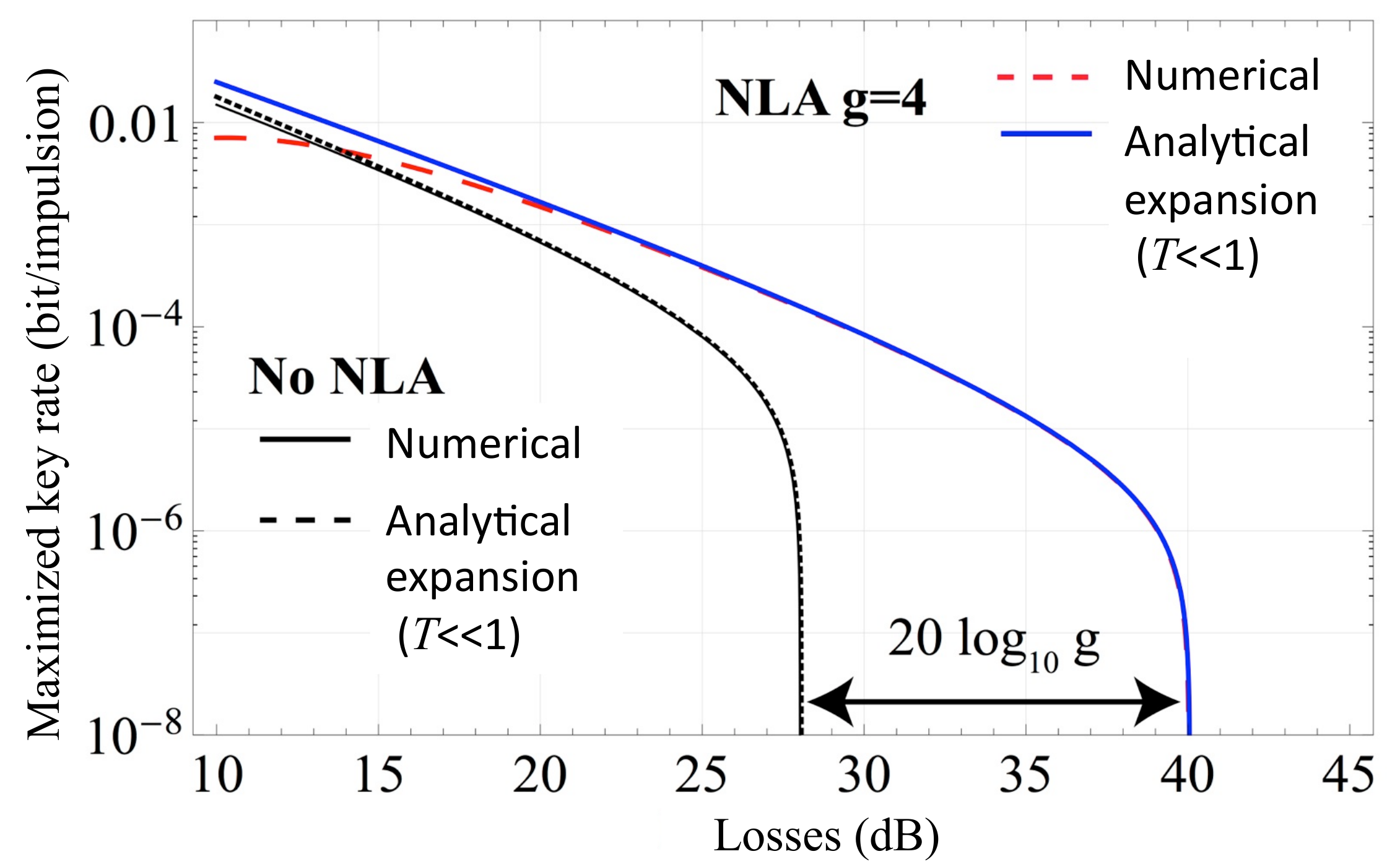} 
\caption{Maximized secret key rate against losses in dB. The maximization is performed on $\lambda$ for the series expansion, and on $\zeta$ for the the numerical expression. The numerical curves are in excellent agreement with the analytical expansions. As explained in the main text, the secret key rate with the NLA is very optimistic due to the probability of success $1/g^2$, and hence its curve gives only information on its positivity. The other parameters are $\epsilon{=}0.05$, $\beta{=}0.95$ \cite{JKL11}.}
\label{schemaEBPM3}
\end{figure}

\begin{figure}[h]
\centering
\includegraphics[width=\columnwidth]{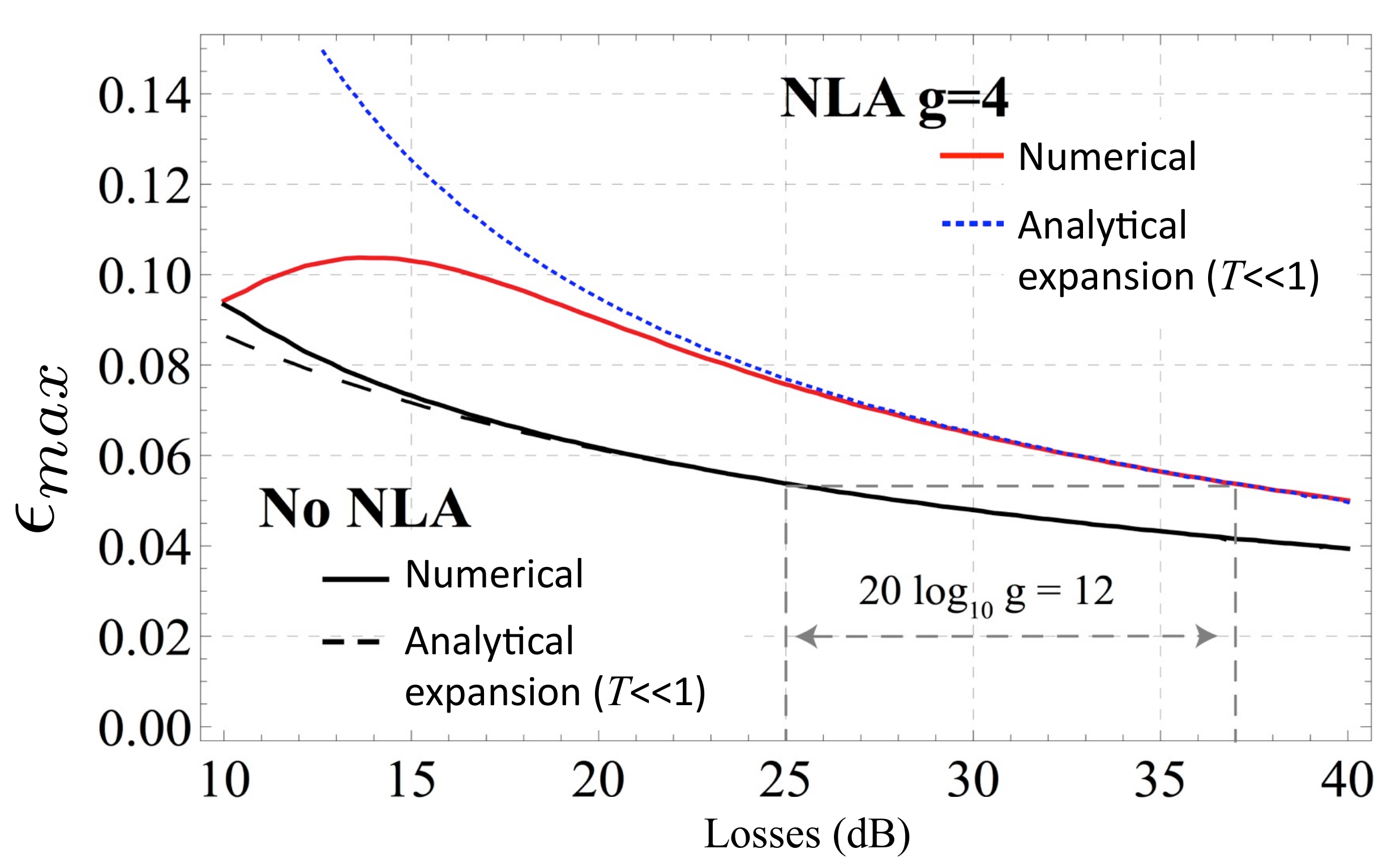} 
\caption{Maximal excess noise for which the secret key rate is positive, against losses. The curves do not depend on the probability of success chosen for the NLA.  The maximization is performed on $\lambda$ for the series expansion, and on $\zeta$ for the numerical expression. For low losses, we see that the first order expansion is not enough, whereas it is in excellent agreement with the numerical curve for strong losses.  The reconciliation efficiency is $\beta{=}0.95$ \cite{JKL11}.}
\label{contourPlot}
\end{figure}

Those results are compared with numerical results for the full expressions of $\Delta I$ and $\Delta I_\mathrm{NLA}$, on Fig. \ref{schemaEBPM3} and \ref{contourPlot}. For both figures, the secret key rate is computed without the NLA and with a NLA of gain $g{=}4$ (which is in the allowed region of Fig. \ref{gMax}). 


 Those figures  clearly show that the secret key rate stays positive for losses increased by $\Delta\mathfrak{L}{=}12$ dB. Fig.  \ref{contourPlot} also shows that for given losses, the secret key rate stays positive for a higher value of excess noise. However, the gain in excess noise depends on the losses, and does not have a simple analytical expression.
  
\begin{figure}[H]
\centering
\includegraphics[width=8cm]{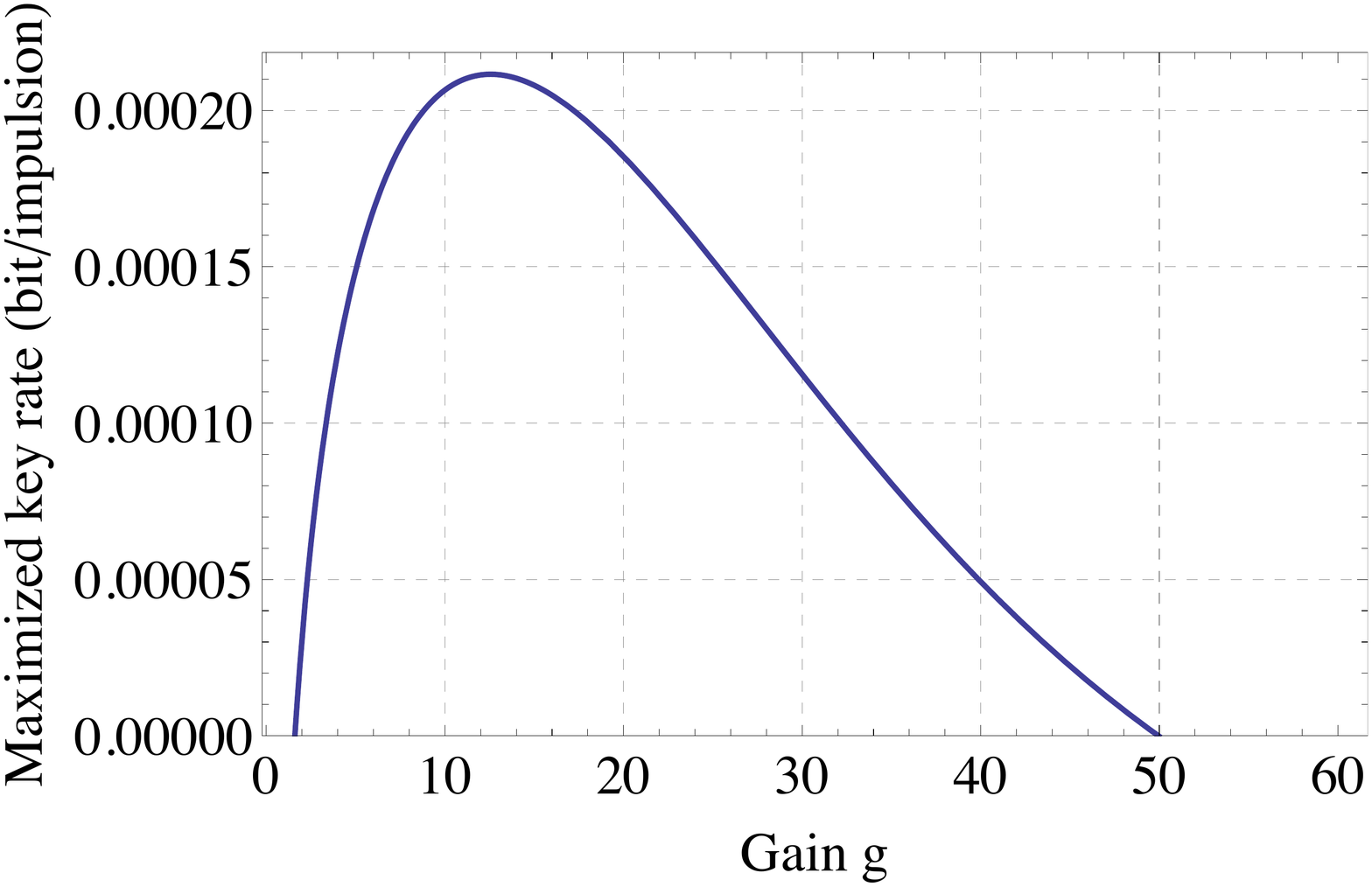} 
\caption{Maximized secret key rate against the gain of the NLA, with a probability of success $1/g^2$. $\beta{=}1$, $\epsilon{=}0.1$, losses${=}$30 dB. With a gain $g{=}1$, the secret key rate is negative. The NLA can increase the secret key rate to positive values when the gain is increased up to a certain value, however if the gain is too important the secret key rate decreases and becomes negative again. The reason is that the effective excess noise becomes too important for the effective transmittance.}
\label{keyRateVSgain}
\end{figure}

 Another important remark concerns the optimal gain. If the transmission can be intuitively increased by increasing the gain, this is not always the case for the secret key rate, as shown on Fig. \ref{keyRateVSgain}. The first reason is the competition between the decreasing probability of success $1/g^2$ and the potential increase of the secret key rate for the successfully amplified states. The second reason is due to the dependance on the gain of the effective parameters (Eq. \ref{parametresEffectifsMain}): the higher the gain, the higher $\eta$, but also the higher $\epsilon^g$. If the gain is too high, it is thus possible that the effective excess noise $\epsilon^g$ would be too important, for the transmittance $\eta$, to give a positive $\Delta I_\mathrm{NLA}$.

  ~\\

\section{Discussion and conclusion}

In presence of excess noise, the secret key rate of the GG02 protocol against Gaussian collective attacks always becomes negative for a certain distance of transmission. We have shown that the noiseless linear amplifier can increase this distance by the equivalent of $20 \log_{10}g$ dB of losses. We have also shown that for given losses, the protocol is more robust against excess noise.

Our calculation of the secret key rate with the amplifier was based on an effective system for which the security proofs are well established. This approach could also find applications in other quantum communication protocols involving an EPR state sent through a quantum channel, followed by a noiseless amplifier. In particular, it could be applied to other CV QKD protocols, for instance protocols using squeezed states or with a heterodyne detection \cite{WLB04,LSS05,WLB06,PhysRevLett.102.130501} 

A further work would be to consider the experimentally demonstrated schemes of the NLA, which are only valid approximations of the ideal NLA up to a certain number of photons. If the state can be well approximated by this truncation, so that the Gaussian approximation still holds, the results presented in this paper are still valid. On the other hand, if the Gaussian approximation does not hold anymore, security proofs are more complicated to manipulate. This problem lies beyond the scope of the present work, and deserves further investigation.

We acknowledge support from the EU project ANR ERA-Net CHISTERA HIPERCOM. MB
is supported by the Marie Curie contract PIEF-GA-2009-236345-PROMETEO.


\appendix


\section{Derivation of the effective parameters}
\label{appendixEffectiveParameters}
In this appendix, we detail the method used to obtain the effective parameters of Section \ref{equivalentChannel}. Let us start by first computing the output of the NLA when the input state $\op{\rho}$ is a thermal state $\op{\rho}_\mathrm{th}(\lambda_\mathrm{ch}){=}(1{-}\lambda_\mathrm{ch}^2)\sum_{n{=}0}^{\infty}\lambda_\mathrm{ch}^{2n}\ket{n}\bra{n}$ displaced by $\beta{=}\beta_\mathrm{x}{+}i\beta_\mathrm{y}$:
\begin{align}
\op{\rho}{=}\op{D}(\beta)\op{\rho}_\mathrm{th}(\lambda_\mathrm{ch})\op{D}({-}\beta) .
\end{align}
This would be the state received by Bob if he knew the result of Alice's heterodyne measurement. The state $\op{\rho}$ can be decomposed on an ensemble of coherent states using the $P$ function:
 \begin{align}
 \op{\rho}=\int P(\alpha)\ket{\alpha}\bra{\alpha} \mathrm{d}\alpha.
 \label{displThermState}
 \end{align}
 where $P(\alpha){=}\frac{e^{\vert\alpha \vert^2}}{\pi^2}\int e^{\vert u \vert^2} \bra{{-}u}\op{\rho}\ket{u}e^{u^*\alpha{-}u \alpha^*} \mathrm{d}u$. Straightforward calculations show that $\bra{{-}u{-}\beta}\op{\rho}_\mathrm{th}(\lambda_\mathrm{ch})\ket{u{-}\beta}=\\ (1{-}\lambda_\mathrm{ch}^2)e^{{-}\vert u \vert^2(1{+}\lambda_\mathrm{ch}^2){-}\vert \beta \vert ^2(1{-}\lambda_\mathrm{ch}^2){+}(u\beta^*{-}u^*\beta)(1{-}\lambda_\mathrm{ch}^2)}$, and therefore  $P(\alpha_\mathrm{x}{+}i\alpha_\mathrm{y}){=}p(\alpha_\mathrm{x}) p(\alpha_\mathrm{y})$, with
\begin{align}
 p(\alpha_\mathrm{x})=\frac{1}{\sqrt{\pi}}\sqrt{\frac{1-\lambda_\mathrm{ch}^2}{\lambda_\mathrm{ch}^2}}e^{-\frac{1{-}\lambda_\mathrm{ch}^2}{\lambda_\mathrm{ch}^2}(\alpha_\mathrm{x}-\beta_\mathrm{x})^2}. 
 \label{pFunc}
 \end{align}
 In the absence of thermal noise ($\lambda_\mathrm{ch}{=}0$), the expression (\ref{pFunc}) becomes proportional to a Dirac distribution $\delta(\alpha_\mathrm{x}{-}\beta_ \mathrm{x})$. The same statements hold for $p(\alpha_\mathrm{y})$. 
 
 The successful amplification can ideally be described by an operator $\op{C}{=}g^{\hat{n}}$,   where $\hat{n}$ is the number operator in the Fock basis. The final state has to be normalized, but one has to be careful that the norm is not the success probability of the transformation, since $\op{C}$ is unbounded. The amplification of a coherent state $\ket{\alpha}$ leads to an amplified coherent state proportional to $\ket{g\alpha}$:
 \begin{align}
  \op{C}\ket{\alpha}{=}e^{\frac{\vert \alpha \vert ^2}{2}(g^2{-}1)}\ket{g\alpha}.
  \label{cohState}
  \end{align}
 Since $\op{C}$ is linear, the amplification of $\op{\rho}$ is simple to derive, using (\ref{pFunc}) and (\ref{cohState}) in  the decomposition (\ref{displThermState}):
\begin{align}
\op{\rho}'&=\op{C}\op{\rho}\op{C}\\
&=\int P(\alpha)e^{\vert \alpha \vert ^2(g^2-1)}\ket{g\alpha}\bra{g\alpha}\mathrm{d}\alpha.
\end{align} 
By introducing the change of variable $u{=}g \alpha$, one gets
\begin{align}
\op{\rho}'=\int P(u/g)e^{\frac{g^2-1}{g^2}\vert u^2 \vert}\ket{u}\bra{u}\mathrm{d}u.
\end{align}
As before, it is easy to see that $P(u/g){=}p(u_\mathrm{x}/g)p(u_\mathrm{y}/g)$. Since $\vert u^2 \vert{=}u_\mathrm{x}^2{+}u_\mathrm{y}^2$, we can consider only the term $p(u_\mathrm{x}/g)\exp\left(\frac{g^2-1}{g^2}u_\mathrm{x}^2\right)$, the results being similar for $u_\mathrm{y}$:
\begin{align}
p(u_\mathrm{x}/g)e^{\frac{g^2{-}1}{g^2}u_\mathrm{x}^2}&{=}\frac{1}{\sqrt{\pi}}\sqrt{\frac{1{-}\lambda_\mathrm{ch}^2}{\lambda_\mathrm{ch}^2}}e^{{-}\frac{1{-}\lambda_\mathrm{ch}^2}{\lambda_\mathrm{ch}^2}(\frac{u_\mathrm{x}}{g}{-}\beta_\mathrm{x})^2+\frac{g^2{-}1}{g^2}u_\mathrm{x}^2}.
\end{align}

The argument of the exponential can easily put in the form:
\begin{align}
&{-}\frac{1{-}\lambda_\mathrm{ch}^2}{\lambda_\mathrm{ch}^2}\Big(\frac{u}{g}{-}\beta_\mathrm{x}\Big)^2{+}\frac{g^2{-}1}{g^2}u^2= \nonumber \\
&\underbrace{{-}\frac{1{-}g^2\lambda_\mathrm{ch}^2}{g^2\lambda_\mathrm{ch}^2}}_{\substack{\text{Thermal state $g\lambda_\mathrm{ch}$}}}\Big(u-\beta_\mathrm{x} \underbrace{g\frac{1{-}\lambda_\mathrm{ch}^2}{1{-}g^2\lambda_\mathrm{ch}^2}}_{\text{Effective gain}}\Big)^2{-}\underbrace{\beta_\mathrm{x}^2\frac{(1{-}g^2)(1{-}\lambda_\mathrm{ch}^2)}{1{-}g^2\lambda_\mathrm{ch}^2}}_{\substack{\text{Normalization term}}}.
\label{pEff}
\end{align}
Thus, the expression (\ref{pEff}) clearly corresponds to a thermal state $\op{\rho}_\mathrm{th}(g\lambda_\mathrm{ch})$ displaced by $g\frac{1-\lambda_\mathrm{ch}^2}{1-g^2\lambda_\mathrm{ch}^2}\beta$, up to a global unimportant normalization factor independent of the variable integrated $\alpha$ or $u$. We can conclude that:
\begin{align}
\op{\rho}'\propto \op{D}(\tilde{g}\beta)\op{\rho}_\mathrm{th}(g\lambda_\mathrm{ch})\op{D}(-\tilde{g}\beta).
\label{NLAdispTherm}
\end{align}
where $\tilde{g}=g\frac{1-\lambda_\mathrm{ch}^2}{1-g^2\lambda_\mathrm{ch}^2}$. In order to keep a physical interpretation, we note that $g$ must be such that $g\lambda_\mathrm{ch}{<}1$.

Let us now find the values of $\beta$ and $\lambda_\mathrm{ch}$ corresponding to the Entanglement-Based protocol presented in the main text. When Alice obtains the results $\alpha_\mathrm{A}$ for her heterodyne measurement on one mode of the EPR state $\ket{\lambda}$, the second mode is projected on a coherent state with an amplitude proportional to $\lambda \alpha_\mathrm{A}$ \cite{grosshans_virtual_2003}. This state is then sent through the quantum channel of transmittance $T$, with transforms its amplitude to $\propto \sqrt{T}\lambda \alpha_\mathrm{A}$. The displacement $\beta$ can thus be taken as:
\begin{align}
\beta=\sqrt{T}\lambda \alpha_\mathrm{A}.
\label{betaDis}
\end{align}
The variance $\frac{1{+}\lambda_\mathrm{ch}^2}{1{-}\lambda_\mathrm{ch}^2}$ of the thermal state corresponds to Bob's variance $1{+}T\epsilon$ when $V_\mathrm{A}{=}0$:
\begin{align}
\frac{1+\lambda_\mathrm{ch}^2}{1-\lambda_\mathrm{ch}^2}=1+T\epsilon \Rightarrow \lambda_\mathrm{ch}^2=\frac{T\epsilon}{2+T\epsilon}.
\label{lambdaCH}
\end{align}
Finally, the action of the NLA (Eq. \ref{NLAdispTherm}) on a displaced thermal state given by Eq. \ref{betaDis} and \ref{lambdaCH} induces the transformations:
\begin{equation} 
	\addtolength{\fboxsep}{4pt}  
				\begin{split} 
					\sqrt{T} \lambda \alpha_\mathrm{A} &\underset{\mathrm{NLA}}{\rightarrow} g\frac{1-\lambda_\mathrm{ch}^2}{1-g^2\lambda_\mathrm{ch}^2} \sqrt{T} \lambda \alpha_\mathrm{A}, \\
					\frac{T\epsilon}{2+T\epsilon} &\underset{\mathrm{NLA}}{\rightarrow} g^2\frac{T\epsilon}{2+T\epsilon}.
					\label{e29}
				\end{split}  
	\end{equation}
The next step is to consider the action of the NLA when Bob does not have any knowledge on Alice's measurement outcome. In such a case, his state is a thermal state $\op{\rho}_\mathrm{B}{=}(1{-}\lambda^{\star2})\sum_{n{=}0}^{\infty}(\lambda^\star)^{2n}\ket{n}\bra{n}$, whose variance is given by $\gamma_\mathrm{AB}$:
\begin{align}
 \frac{1{+}\lambda^{\star2}}{1{-}\lambda^{\star2}}=1{+}T V_\mathrm{A}{+}T \epsilon {\Rightarrow} \lambda^{\star2}{=}\frac{T \left(\lambda ^2 \left(2{-}\epsilon\right){+}\epsilon\right)}{2{-}\lambda ^2 \left(2{+}T\left(\epsilon{-}2\right)\right){+}T \epsilon} .
 \label{e31}
 \end{align}
 Since the NLA always transforms a thermal state of parameter $\lambda^{\star}$ into another thermal state of parameter $g\lambda^{\star}$, Eq. \ref{e31} shows that the NLA performs the transformation:
 \begin{align}
\frac{T \left(\lambda ^2 \left(2{-}\epsilon\right){+}\epsilon\right)}{2{-}\lambda ^2 \left(2{+}T\left(\epsilon{-}2\right)\right){+}T \epsilon} \underset{\mathrm{NLA}}{\rightarrow} g^2 \frac{T \left(\lambda ^2 \left(2{-}\epsilon\right){+}\epsilon\right)}{2{-}\lambda ^2 \left(2{+}T\left(\epsilon{-}2\right)\right){+}T \epsilon}.
\label{e32}
\end{align}

We have now all the required equations to find the expression of the effective parameters $\zeta$, $\eta$ and $\epsilon^g$. Using Eq. \ref{e29} and \ref{e32}, those parameters must satisfy:
\begin{align}
\begin{split}
\sqrt{\eta} \zeta  &{=} g\frac{1{-}\lambda_\mathrm{ch}^2}{1{-}g^2\lambda_\mathrm{ch}^2} \sqrt{T} \lambda,  \\
\frac{\eta\epsilon^g}{2{+}\eta\epsilon^g} &{=} g^2\frac{T\epsilon}{2{+}T\epsilon},\\
\frac{\eta \left(\zeta ^2 \left(2{-}\epsilon^g\right){+}\epsilon^g\right)}{2{-}\zeta ^2 \left(2+\eta\left(\epsilon^g{-}2\right)\right){+}\eta \epsilon^g} &{=} g^2 \frac{T \left(\lambda ^2 \left(2{-}\epsilon\right){+}\epsilon\right)}{2{-}\lambda ^2 \left(2{+}T\left(\epsilon{-}2\right)\right){+}T \epsilon}.
\end{split}
\end{align}
This system can be solved, leading to:
\begin{equation} 
	\addtolength{\fboxsep}{4pt} 
				\begin{split} 
					\zeta&=\lambda  \sqrt{\frac{\left(g^2-1\right) \left(\epsilon-2\right) T-2}{\left(g^2-1\right)
   \epsilon T-2}} ,\\
\eta&=\frac{g^2 T}{\left(g^2-1\right) T \left(\frac{1}{4} \left(g^2-1\right) \left(\epsilon-2\right) \epsilon T-\epsilon+1\right)+1} , \\
\epsilon^g&=\epsilon-\frac{1}{2} \left(g^2-1\right) \left(\epsilon-2\right) \epsilon T.
					\label{parametresEffectifs}
				\end{split}  
	\end{equation}

Finally, the expression of the maximum gain $g_\mathrm{max}(T,\epsilon)$ for which those parameters take physical values is given by:

\begin{align}
&g_\mathrm{max}(T,\epsilon)=\\
&\sqrt{\frac{\epsilon \left(T \left(\epsilon{-}4\right){+}2\right){+}4 \sqrt{\frac{T \left(\epsilon-2\right){+}2}{\epsilon}}{-}2 \sqrt{\epsilon \left(T \left(\epsilon{-}2\right){+}2\right)}{+}4
   T{-}4}{T \left(\epsilon{-}2\right){}^2}}.
   \label{gainMax}
   \end{align}
~\\
Let us stress some important comments about those effective parameters, which confirm the validity of their expression.  First, they naturally reduce to the real physical parameters without the NLA, for $g{=}1$:
\begin{align}
g=1 \Rightarrow \left\{ 
	\begin{array}{l}
\zeta=\lambda\\
\eta=T\\
\epsilon^g= \epsilon
	\end{array} \right.
\end{align}
Then, when there is no excess noise ($\epsilon{=}0$), they match previous results \cite{ralph_nondeterministic_2008}:
\begin{align}
\epsilon=0 \Rightarrow \left\{ 
	\begin{array}{l}
\zeta=\lambda \sqrt{1+(g^2-1)T} \\
\eta=\frac{g^2T}{1+(g^2-1)T}\\
\epsilon^g= 0
	\end{array} \right.
\end{align}

\begin{widetext}
\section{Expressions used to compute the Holevo bound $\chi_{BE}$}
\label{appendixHolevo}

The Holevo bound $\chi_\mathrm{BE}$ is given by \cite{LBG07}:
 $\chi_\mathrm{BE}=G\left[ \frac{\mu_1-1}{2} \right]+G\left[ \frac{\mu_2-1}{2} \right]-G\left[ \frac{\mu_3-1}{2} \right]-G\left[ \frac{\mu_4-1}{2} \right]
$
where
\begin{align}
G[x]&=(x+1)\log_2 [x+1]-x\log_2[x] \text{ si x $\neq$0, et } G[0]=0 \\
\mu_{1,2}^2&=\frac{1}{2}\left(A \pm \sqrt{A^2-4E}\right) \qquad  \mu_{3,4}^2=\frac{1}{2}\left(C \pm \sqrt{C^2-4D}\right) \\
A&=V^2(1-2T)+2 T +T^2(V+\chi_\mathrm{line})^2 \qquad E=T^2(V \chi_\mathrm{line}+1)^2 \\
C&=\frac{V \sqrt{E}+T(V+\chi_\mathrm{line})}{T(V+\chi_\mathrm{line})} \qquad D=\frac{\sqrt{E}V}{T(V+\chi_\mathrm{line})}\\
\end{align}  
$V{=}V_\mathrm{A}{+}1$ is the variance of Alice's thermal state (see text for details), and $\chi_\mathrm{line}=\frac{1-T}{T}+\epsilon$ is the total equivalent input noise. Bob's homodyne detection is assumed to be perfect.

\section{First order expansion in T}
\label{appendixDLHolevo}
The first order expansion in T of the secret key rate given in Appendix \ref{appendixHolevo} using the NLA is:
\begin{align}
&\Delta I_\mathrm{NLA} {\simeq} \nonumber\\ 
&P_\mathrm{suc} g^2 T\frac{ (-2 \beta \lambda^2 (-1 {+} \lambda^2) - \epsilon (-1 {+} \lambda^2)^2 (1 {+} 
      \log[2]) {+} (-1 {+} \lambda^2) (\epsilon (-1 {+} \lambda^2)(\log[ \epsilon] {+} \log[g^2 T]) {+} 
      4 \lambda^2 \log[\lambda]) {+} 
   2 \lambda^2 \log[\lambda^2]) }{(2 (-1 {+} \lambda^2)^2 \log[2]}.
   \label{dlHolevo}
\end{align}

The equation that must satisfy the optimal value $\lambda_\mathrm{opt}$ to maximize the transmission distance (Eq. \ref{Tlim}), and maximize the secret key rate (Eq. \ref{dlHolevo}), is given by :
\begin{align}
\frac{\lambda_\mathrm{opt} ^2 \left(\lambda_\mathrm{opt} ^2-4 \log (\lambda_\mathrm{opt} )-1\right)}{1-\lambda_\mathrm{opt} ^2}=\beta.
\label{betaOpt}
\end{align}

Fig. \ref{betaVSLambdaOpt} shows $\lambda_\mathrm{opt}$ as a function of $\beta$:

 \begin{figure}[H]
\centering
\includegraphics[width=8cm]{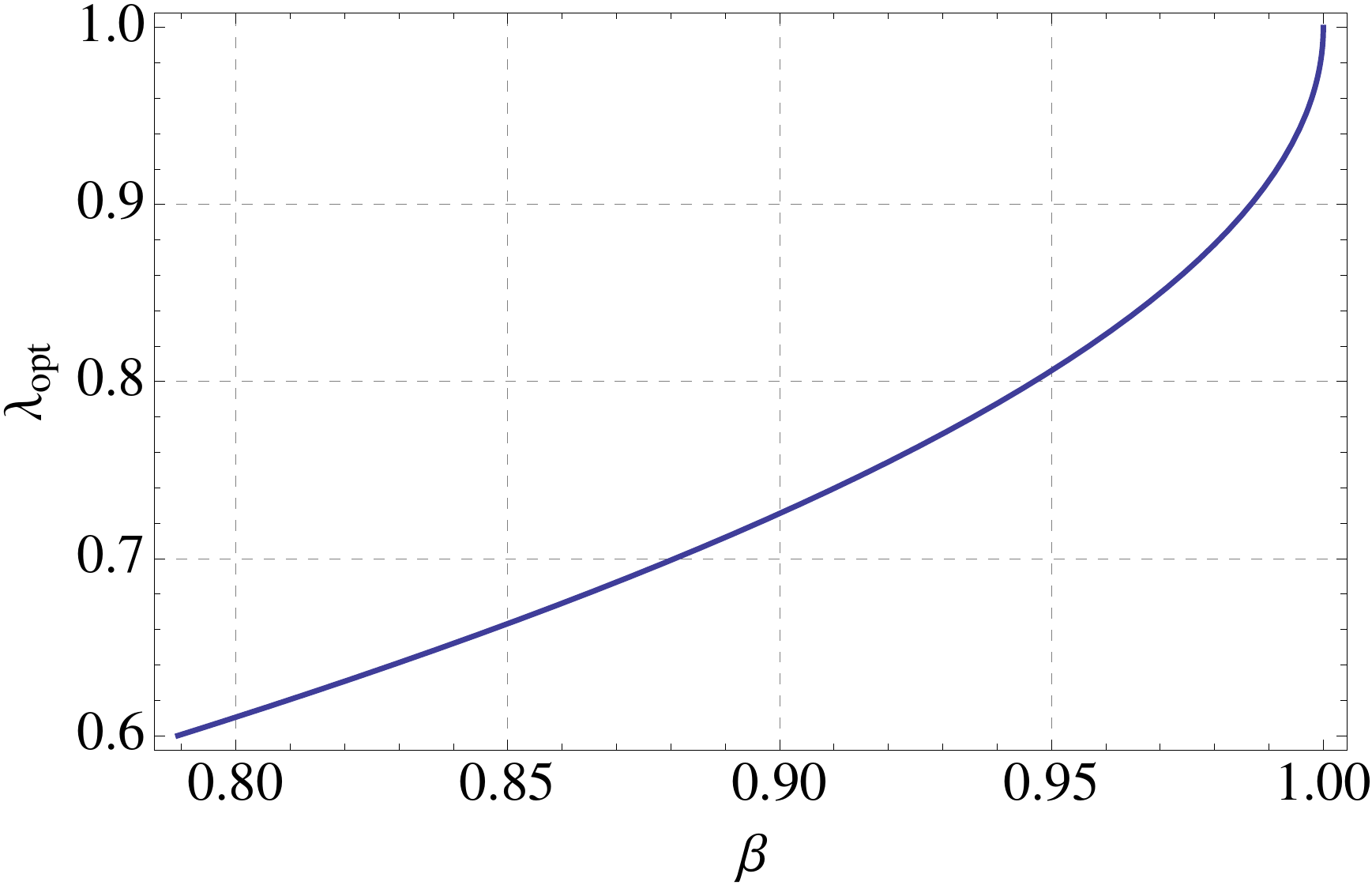} 
\caption{$\lambda_\mathrm{opt}$ against $\beta$. $\beta{=}0.95$ gives $\lambda_\mathrm{opt}{\simeq}0.806$.}
\label{betaVSLambdaOpt}
\end{figure}

\end{widetext}

\section{Success probability}
\label{appendixProba}
The success probability of the NLA can depend on many experimental factors. Here, we are interested in deriving an upper bound based on very general principles, when the success probability can be considered as being a constant value. In this way, we can obtain an optimistic estimate of its performance, but certainly we will not overlook interesting regimes. In both EB and PM versions of the GG02 protocol, Bob's state prior to any classical communication with Alice is the thermal state $\op{\rho}_\mathrm{B}(\lambda^{\star}){=}(1{-}\lambda^{\star2})\sum_{n{=}0}^{\infty}(\lambda^\star)^{2n}\ket{n}\bra{n}$.

 Consider now that the NLA produces an amplified state $\op{\rho}_\mathrm{B}(g\lambda^{\star})$ with a success probability $P_\mathrm{suc}$. When the amplification fails, the protocol is aborted, and the state is simply replaced by the vacuum $\ket{0}\bra{0}$. Without post-selection, the NLA can therefore be represented as a trace preserving operation $\mathcal{T}$ described by:
 \begin{align}
 \mathcal{T}( \op{\rho}_\mathrm{B}(\lambda^{\star}))=P_\mathrm{suc} \op{\rho}_\mathrm{B}(g\lambda^{\star})+(1-P_\mathrm{suc})\ket{0}\bra{0}.
 \label{NLAT}
 \end{align}
 Naturally,  $\mathcal{T}$ applied on the vacuum also gives the vacuum, regardless of the value of $P_\mathrm{suc}$. Since any trace preserving quantum operation cannot decrease the fidelity $\mathcal{F}$ between two quantum states \cite{nielsen_quantum_2000}, $\mathcal{T}$ must verify:
 \begin{align}
\mathcal{F} \Big(\op{\rho}_\mathrm{B}(\lambda^{\star}), \ket{0}\bra{0}\Big)&\leq \mathcal{F} \Big(\mathcal{T}(\op{\rho}_\mathrm{B}(g\lambda^{\star}), \ket{0}\bra{0}\Big).
\label{bornFid0}
\end{align}
 which gives us an upper bound on $P_\mathrm{suc}$. Indeed, inserting the expression for Bob's transformed state (Eq. \ref{NLAT}) in the constraints on fidelities (Eq. \ref{bornFid0}), we find that
 $P_\mathrm{suc}$ must verify:
\begin{align}
\bra{0} \op{\rho}_\mathrm{B}(\lambda^{\star}) \ket{0}&\leq \bra{0}\Big(P_\mathrm{suc}\op{\rho}_\mathrm{B}(g\lambda^{\star}){+}(1{-}P_\mathrm{suc})\ketbra{0}{0}\Big)\ket{0} .
\label{bornFid}
\end{align}
which is satisfied if:
\begin{align}
P_\mathrm{suc}\leq\frac{1}{g^2}.
\end{align}

\end{document}